# DALiuGE: A Graph Execution Framework for Harnessing the Astronomical Data Deluge

Chen Wu[a,*], Rodrigo Tobar[a], Kevin Vinsen[a], Andreas Wicenec[a], Dave Pallot[a],
Baoqiang Lao[b], Ruonan Wang[c], Tao An[b], Mark Boulton[a], Ian Cooper[a], Richard Dodson[a],
Markus Dolensky[a], Ying Mei[d,e], Feng Wang[d,e]

[a] *International Centre for Radio Astronomy Research (ICRAR),*
*The University of Western Australia,*
*M468, 35 Stirling Highway, Crawley, Perth, WA 6009, Australia*
[b] *Shanghai Astronomical Observatory,*
*Shanghai, China*
[c] *Oak Ridge National Laboratory,*
*TN 37831, United States*
[d] *Kunming University of Science and Technology,*
*Kunming, Yunnan, China*
[e] *Yunnan Observatories, Chinese Academy of Sciences,*
*Kunming, Yunnan, China*

**Abstract**

The Data Activated *Liu*[1] Graph Engine — DALiuGE[2] — is an execution framework for processing large astronomical datasets at a scale required by the Square Kilometre Array Phase 1 (SKA1). It includes an interface for expressing complex data reduction pipelines consisting of both data sets and algorithmic components and an implementation run-time to execute such pipelines on distributed resources. By mapping the logical view of a pipeline to its physical realisation, DALiuGE separates the concerns of multiple stakeholders, allowing them to collectively optimise large-scale data processing solutions in a coherent manner. The execution in DALiuGE is data-activated, where each individual data item autonomously triggers the processing on itself. Such decentralisation also makes the execution framework very scalable and flexible, supporting pipeline sizes ranging from less than ten tasks running on a laptop to tens of millions of concurrent tasks on the second fastest supercomputer in the world. DALiuGE has been used in production for reducing interferometry data sets from the Karl E. Jansky Very Large Array and the Mingantu Ultrawide Spectral Radioheliograph; and is being developed as the execution framework prototype for the Science Data Processor (SDP) consortium of the Square Kilometre Array (SKA) telescope. This paper presents a technical overview of DALiuGE and discusses case studies from the CHILES and MUSER projects that use DALiuGE to execute production pipelines. In a companion paper, we provide in-depth analysis of DALiuGE's scalability to very large numbers of tasks on two supercomputing facilities.

---

[1] *Liu* is the phonetic form of the Chinese character 流, meaning "flow".
[2] Pronounced [da ˈˡliu̯ ˌɡi] or approximately `Da-'Lieu-Gee`

# 1. Introduction

The Square Kilometre Array (SKA) [1, 2] will be the largest radio telescope in the world. It is to be built from 2018, making it the latest large-scale global scientific endeavour. The first phase of the project — SKA1 [3] — will consist of hundreds of dishes and hundreds of thousands of antennas, enabling the monitoring and surveying of the sky in unprecedented detail and speed, with a second phase expanding these capabilities to at least an order of magnitude. Because of its immense size, just one SKA1 science project will produce correlated data at a rate of 466 GiBytes/s [4] for the low frequency component (SKA1-Low) and 446 GiBytes/s [5] for the mid frequency component (SKA1-Mid). This correlated interferometry data will be fed into the Science Data Processor (SDP) [6, 7], a Many-Task Computing (MTC) [8, 9] centre, responsible for processing and reducing the data; and producing and preserving science-ready products continuously. The SKA1 will have constrained power allocations [10] to process observations as they are performed in real time. This poses considerable challenges to manage, process and store such large datasets.

Traditional High Performance Computing (HPC) facilities are optimised for processing compute-intensive scientific jobs [8, 9]. Therefore they are not well suited to cope with data-intensive workloads, which involve continuous data capturing, processing, analysis, and curation. High Throughput Computing (HTC) on the other hand is concerned with longer term provisioning of large amounts of computing, with the primary metric being operations per month. As a superset of both HPC and HTC [9], MTC combines the stability and dependent tasks of HTC with the very high number of individual independent tasks of HPC. In MTC individual tasks may be small or large, uniprocessor or multiprocessor, compute-intensive or data-intensive; and that is what we need for a typical radio astronomy workflow as well. In addition, these activities are often interleaved with one another inside a sophisticated workflow, requiring highly responsive and data-location aware resources provisioned on demand. Our experience of operating the data system [11] for the SKA-low precursor telescope (the Murchison Widefield Array [12]) suggests that the overheads associated with data migration, access, distribution, and conversion between different formats are increasingly dominating the overall pipeline execution costs. Moreover, the complexity of radio astronomy processing originates not only from the basic algorithmic components (e.g. FFT,

---

*Corresponding Author

*Email addresses:* `chen.wu@icrar.org` (Chen Wu), `rtobar@icrar.org` (Rodrigo Tobar), `kevin.vinsen@icrar.org` (Kevin Vinsen), `andreas.wicenec@icrar.org` (Andreas Wicenec), `dave.pallot@icrar.org` (Dave Pallot), `lbq@shao.ac.cn` (Baoqiang Lao), `wangr1@ornl.gov` (Ruonan Wang), `antao@shao.ac.cn` (Tao An), `mark.boulton@icrar.org` (Mark Boulton), `ian.cooper@icrar.org` (Ian Cooper), `richard.dodson@icrar.org` (Richard Dodson), `markus.dolensky@icrar.org` (Markus Dolensky), `meiying@cnlab.net` (Ying Mei), `cnwangfeng@gmail.com` (Feng Wang)

Gridding, Deconvolution, etc.) but also the complex combinations of ways these components access their input, output, metadata and intermediate data. This makes it very difficult to apply a "one-size-fits-all" strategy (e.g. data re-organisation, I/O overlapping, intelligent caching, etc.) to achieve a global optimum across multiple pipeline stages on distributed HPC resources. Since the current state-of-the-art astronomy data processing systems are designed to handle data approximately two to three orders of magnitude smaller than the SKA1 [4, 5], a new data execution framework is much needed.

Currently the most popular way of processing astronomy data is to define workflow components statically in scripts. These scripts are then either executed sequentially on a local machine or wrapped into job scripts submitted to job scheduling systems such as PBS or SLURM in an HPC environment. Such application-driven workflow models have several drawbacks. Firstly, most astronomical projects are international, involving multiple institutes across the globe. As a result, astronomers often need to tailor or even re-develop workflow scripts in order to make them work — compiling, deploying, running, monitoring, etc. — on computers or clusters with different system setup and hardware architecture. This happens whenever there is an upgrade on the workflow, hardware, telescope configuration, leading to considerable cost. Secondly, the execution process lacks real-time monitoring and control. For example, in many cases users cannot easily determine the status (e.g. success, failure, etc.) of the pipeline execution until the entire workflow is completed or a considerable amount of computing and storage resources have been consumed. For SKA-scale data processing (with tens of millions of concurrent tasks) this is not only infeasible, but it is extremely expensive to delay fault detection and subsequent recovery actions (e.g. re-execution). Finally, even if failures or exceptions are noticed at an earlier stage users still have to restart the entire job for re-execution. This is because a workflow driven by "processing" rather than "data" cannot adjust task execution dynamically based on whatever intermediate datasets (e.g. from previous runs) and resources happen to be available. More "intelligent" workflow management often requires substantial ad-hoc human interactions, which become impractical for SKA-scale data processing workflows.

To tackle the above challenges, we have developed the Data Activated Liu (Flow) Graph Execution Engine — DALiuGE (DA流GE) — for the SDP Consortium of the SKA1 design. DALiuGE aims to provide a distributed data management platform and a scalable pipeline execution environment to support continuous, time and power bounded, data-intensive processing for producing SKA science-ready products. The distinct advantages of DALiuGE over existing processing frameworks are:

1. It explicitly decouples the logical view of a problem from its runtime realisation. This not only separates the concerns of different stakeholders such as: telescope operators, pipeline developers, and astronomers; more importantly, it allows them to collectively optimise data processing at multiple layers in a homogeneous manner, whilst letting the framework optimise the generation of execution plans using resources and profiling information.
2. It is based on, but extends, the dataflow programming paradigm. Unlike traditional dataflow models that characterise data as tokens moving across directed edges between nodes, we instead model data as nodes of the graph, elevating them to manageable entities which are implemented as active objects in memory and can be monitored and persisted if necessary. Continuous streams of data (a central idea of the dataflow paradigm) are handled by the framework by including streaming-oriented applications in the design. In our graph model these generalised nodes are called **Drops** and come in two concrete forms, **Application Drops** (tasks) and **Data Drops** (data).
3. The Drop concept allows data as well as applications to trigger and receive events, which are the tokens travelling through the graph edges. Events activate the cascaded execution of parallel processing tasks. This fulfils the idea of "data-activated" execution, where every Data Drop determines whether and when to trigger the following processing tasks. This completely decentralises the execution orchestration, making the framework scalable to a large number (at least tens of millions) of nodes.
4. It integrates a data lifecycle management component within the execution engine, keeping track of Drops and migrating or deleting them automatically when necessary.

The rest of this paper is organised as follows: Section 2 provides a brief overview of the related work in the area of large-scale data processing. Section 3 introduces key concepts in DALiuGE and their implementation details. A more detailed discussion on the concept of Drop is given in Section 4. Sections 5 and 6 describe real-world use cases of DALiuGE in the CHILES and the MUSER projects respectively, and finally Section 7 concludes the paper based on the current state of DALiuGE and discusses future work.

In a companion paper [13], we evaluate the scalability of DALiuGE — analysing its efficiency against the growth from both the graph size and the resource provision.

## 2. Related work

In this section, we provide a succinct overview of existing work on distributed, large scale data processing frameworks and methods. Given the massive and rapid development in this area, it is infeasible to enumerate all related technologies. Instead, we focus on three main areas, from which we have drawn inspiration to develop DALiuGE. In particular, we identify similarities and compare differences between these techniques and DALiuGE.

*2.1. HPC programming models*

Most HPC applications today use both shared-memory (e.g. OpenMP) and distributed memory (e.g. MPI) programming models to achieve a desirable level of parallelism in the field of high performance computing. Both programming models require developers to exploit potential parallelism, control process/thread synchronisation, and manage various overheads associated with parallelism such as race conditions. However, such extra programming and optimisation effort will be a substantial burden for researchers, difficult for other scientists to maintain and reuse within the community, and are restricted to run on a particular hardware architecture or supercomputer facility. In fact, it has been noted [14] that the uptake of HPC depends on how easy the end-user application software is to use without having to gain expertise in parallel programming. Furthermore, by explicitly engaging in MPI message exchange and collective calls, most MPI applications inevitably separate computation and communication, giving away many potential opportunities for parallelism. Although techniques such as the one-sided MPI communication [15] and strategies [16] to overlap communication and computation alleviate this issue to some degree, they still require dedicated effort and expertise in order to derive embarrassingly parallel solutions.

To ease the development of parallel scientific applications, the Swift/T parallel scripting language [17] allows developers to write C-like functions and expressions using high-level data structures (such as associative arrays). The Swift/T scripts are then compiled into MPI programs before the Turbine dataflow engine [18] schedules and executes them while ensuring dynamic task load balancing among hardware resources. Although Swift/T reduces the complexity of developing compute-intensive applications, it relies on a global file system shared by each worker node to perform file I/O operations. Szalay and Blakeley [19], Dodson et al. [20], and Zhao et al. [21] all discuss how a global file system with its ever increasing large network storage systems and a central metadata server is simply not economically viable and scalable to support data-intensive workloads. In contrast, DALiuGE gives the flexibility to developers to perform I/O in whichever way they deem optimal and productive without relying on a particular storage system architecture.

To tackle the "productivity wall" challenge, the *ClusterSs* programming model [22] allows developers to parallelise the execution of a sequential application across a large cluster without explicitly concerning themselves with distributed data management and flow control. By utilising the APGAS run-time (X10 [23]) as its underlying communication middleware, *ClusterSs* replaces local Java methods defined in the sequential program with dynamically-spawned tasks run by threads on remote nodes, thereby transparently transforming a shared memory-like program into a distributed memory application on-the-fly. DALiuGE also performs similar transformations known as "graph translation". However, unlike *ClusterSs*, such a transformation takes place well before the graph execution starts. As a result, the execution is completely decentralised, no central "main node" is needed in DALiuGE to spawn tasks at run-time, thus significantly reducing framework overheads.

In the past decade, scientific workflow systems and HPC schedulers such as Pegasus [24], Condor [25], and Apache OODT [26] have achieved significant progress for mapping, scheduling, and executing workflows on highly distributed Grid and Cloud resources. Despite their success in handling data-intensive workloads, their primary processing model is task-centric. For example, in the absence of global file systems, the Pegasus `Mapper` explicitly stages data transfers to satisfy dependencies between tasks running on different nodes. DALiuGE, on the other hand, let a Data Drop possess the full knowledge (events, status, operations, etc.) of its payload and trigger processing on itself in a decentralised, autonomous fashion.

*2.2. Data parallel frameworks*

Industrial data-intensive applications often use data parallel frameworks such as MapReduce [27], Dryad [28] or Spark [29] to handle parallel chunks of data independently. Such data parallelism virtually allows processing capabilities to scale indefinitely. While these data parallel frameworks provide horizontally scalable solutions, two issues arise when directly using them out of the box for modelling, executing, and managing (radio) astronomical data processing.

Firstly, data parallelism works by partitioning a large/huge data set (or data stream) into smaller splits (or streams), each of which is then processed in parallel. Most data parallel frameworks are responsible for partitioning, merging, and distributing data splits across distributed hardware resources for optimal performance, load balancing and data parallelism. For example, in the Hadoop MapReduce implementation [30], each file ingested into the Hadoop Filesystem (HDFS) is partitioned by the framework into multiple 64 Megabyte chunks by default. In Spark, an in-memory data frame is partitioned into multiple sub-arrays, each of which is then distributed onto a Worker/Executor for processing. However, partitioning astronomical datasets often involves far more than simply slicing datasets along some dimensions. For example, in our previous work on CHILES data reduction [20], splitting visibility data sets in the frequency domain involves a substantial amount of calculation including FFTs and adjusting Doppler shifts. Whether this kind of complexity is required for the SKA is mainly dependent on the science use case and the array configuration in use. Moreover, different data sets have different partition tolerance and sensitivity. Tailoring a generic, off-the-shelf data parallel framework to account for such science-related nuances would force the intrusion of domain knowledge into the framework itself and that is not desirable.

Secondly, time-critical, deadline-sensitive workloads (like those of the SDP) require guaranteed Quality of Services (QoS, such as execution latency) that are not well supported by existing data parallel frameworks [31]. Although there has been considerable effort in optimising dataflows to meet QoS requirements, the parameter space explored by these efforts is quite different from the context in which radio astronomical data sets are processed. For example, most optimisation strategies during the MapReduce shuffling stage assume that the `Reducer` function supports partial aggregations [32] (for example, the `Combiner` task in the word-count example computes a partial count for each local key). However, many radio astronomical operations (such as CLEAN, FLAGGING, and CALIBRATE) are not commutative and associative, and will produce undesirable artefacts if they are executed as such. On the other hand, optimisation techniques that estimate or predict data and task properties [33, 34] could be very effective for commercial workloads. But their relevance and usefulness in astronomical data pipelines is very limited since the properties (shape, dimension, resolution, etc.) of radio astronomical data sets are precisely defined before and after any data operations in a given pipeline.

To address these two issues the DALiuGE data partitioning and aggregation are modelled as "Scatter" and "Gather" constructs, respectively by the pipeline developer, who explicitly specifies dedicated tasks and defines the relevant partitioning and aggregation parameters based on specific science cases. DALiuGE then uses these parameters to generate optimal dataflows that minimise goals like data movement, execution latency, resource footprint, etc. The separation between the logical graph and the physical dataflow, and the optimal translation between the two allow DALiuGE to achieve "science-compliant" data parallelism while meeting performance QoS constraints.

*2.3. Dataflow and Graph*

The *dataflow* computation model was initially proposed [35, 36, 37] to express programs as Directed Acyclic Graphs (DAG), where the vertices are the stateless computational tasks that compose the program, and edges connect the output of one task with the input of another. Travelling through the edges are data tokens, which are continuously processed by the nodes as they arrive. Unlike OpenMP or MPI, this paradigm does not explicitly place any control or constraints on the order or timing of operations beyond what is inherent in the data dependencies among compute tasks, and therefore exploits the full inherent parallelism of a program. Putting the concept of dataflow into practice still requires control flow operations and data storage in order to make it practical and useful (e.g. the MIT dataflow architecture [38]).

More recently, graph-based models have been used for building scalable data processing pipelines. For example, Luigi [39] was originally developed at Spotify to run thousands of data processing tasks every day, all of which are organised in complex dependency graphs. As an open source project, Luigi has been used by companies to run "multi-stage" data pipelines in production [40]. The Luigi Python APIs allow developers to focus solely on defining individual tasks, while the Luigi scheduler is responsible for building the "global" task dependency graph and assigning tasks to workers for execution. Compared to DALiuGE, which also composes event dependency graphs from individual Drops, Luigi lacks the aforementioned benefits of separation between a logical graph and its physical dataflow realisation. Moreover, the Luigi central scheduler is a scalability bottleneck when executing thousands of tasks in a graph [41]. In contrast, DALiuGE can concurrently manage and execute tens of millions of tasks thanks to its completely decentralised execution model.

TensorFlow [42], which is being developed by the Google Brain team, aims to tackle the processing challenge associated with deep learning — for example, training machine learning models (e.g. deep neural networks with hundreds of billions of parameters) on large scale clusters or perform model inference on mobile platforms. At its core, TensorFlow (1) represents both computations and states (that the computation operates on) as dataflow-like graphs and; (2) maps graphs onto various hardware platforms such as: laptops, GPU clusters, Android/iOS devices, etc.

Although both TensorFlow and DALiuGE treat computation as graphs, and stress the need for "mapping" of dataflow graphs to hardware resources, they differ in two significant ways. First, TensorFlow does not strictly separate logical graphs from physical graphs. For example, when building a Tensorflow graph, a developer can wrap logical computation expressions with concrete physical resource information (e.g. an IP address with a device identifier), thereby enforcing where the computation should take place during run-time. However, we believe a completely resource-oblivious logical graph has more advantage for optimising long-running, large-scale pipelines, and such separation of concerns is vital to astronomical projects with multiple stakeholders.

Second, a TensorFlow graph allows mutable graph nodes — *Variables* — to hold persistent data that can be continuously updated. However in DALiuGE, the payload of data Drops is strictly "write once, read many" as in the traditional dataflow model. This led to very different approaches to realising iterative algorithms essential for both astronomi-cal data reduction (e.g. image reconstruction [43, 44]) and machine learning optimisation (e.g. gradient descent [45]). A TensorFlow graph does not contain loops internally but is repeatedly "run" by an external `Session` object to perform loop iterations during run-time. Only *Variables* in the graph are persistent — their states and data span multiple executions of the same graph. In comparison, a DALiuGE graph contains pre-generated loop structures with new Data Drops created in each iteration. Although a TensorFlow session gives users more flexibility at run-time (e.g. to control loop conditions), DALiuGE is dedicated to guaranteed, optimal resource provisioning, which is important to support real-time data reduction pipelines in large astronomical observatories like the SKA.

## 3. Key concepts and implementation

Setting up a pipeline in DALiuGE typically goes through the following six stages (see Figure 1), all of which are related to the concept of graphs.

1. Logical Graph Component Development,
2. Logical Graph Template Composition,
3. Logical Graph Selection,
4. Physical Graph Template Translation,
5. Physical Graph Deployment, and
6. Physical Graph Execution

This clear enumeration of stages is crucial, as it explicitly separates the different concerns involved in the design and execution of a pipeline, spreading them across both time and domain, allowing different users and stakeholders of the system to concentrate on their domain-specific expertise. Although these stages are defined within the context of the SKA SDP, they are generally applicable to the development, deployment, and execution of many astronomical processing pipelines.

**Stage 1.** *Pipeline Components* are astronomical computational tasks wrapped into ApplicationDrops. Currently such tasks can be implemented as Docker images [46], binary executables, shell scripts or python modules. This flexible wrapping allows DALiuGE to re-use a lot of existing code, without any modification. Each pipeline component is parametrised by *Resource Usage* information measured by the *Resource Analyser*. Pipeline components have to be developed and optimised for the actual target platform. While the computational tasks are stateless, the Application Drops are stateful, since they go through the Drop lifecycle (Figure 11) — to be initialised, deployed, expired and removed, depending on the current state of the overall process.

**Stage 2.** A staff astronomer composes a *Logical Graph Template* representing a high-level data processing capability (e.g., "Image visibility data") using resource-oblivious dataflow constructs and Pipeline Components that have been tested and released. A Logical Graph Template leaves some room for the principle investigators (PIs) requesting an observation to adjust the processing to meet their science goals. It will be possible, for example, to specify the number of output channels for spectral line and continuum data products. A Logical Graph Template in general represents a quite complex data and component orchestration specification and will require careful testing and verification before it is offered as a capability to the PIs. The set of released Logical Graph Template will reside in a fully version and configuration controlled repository and essentially define the various operation modes of the SKA Science Data Processor.

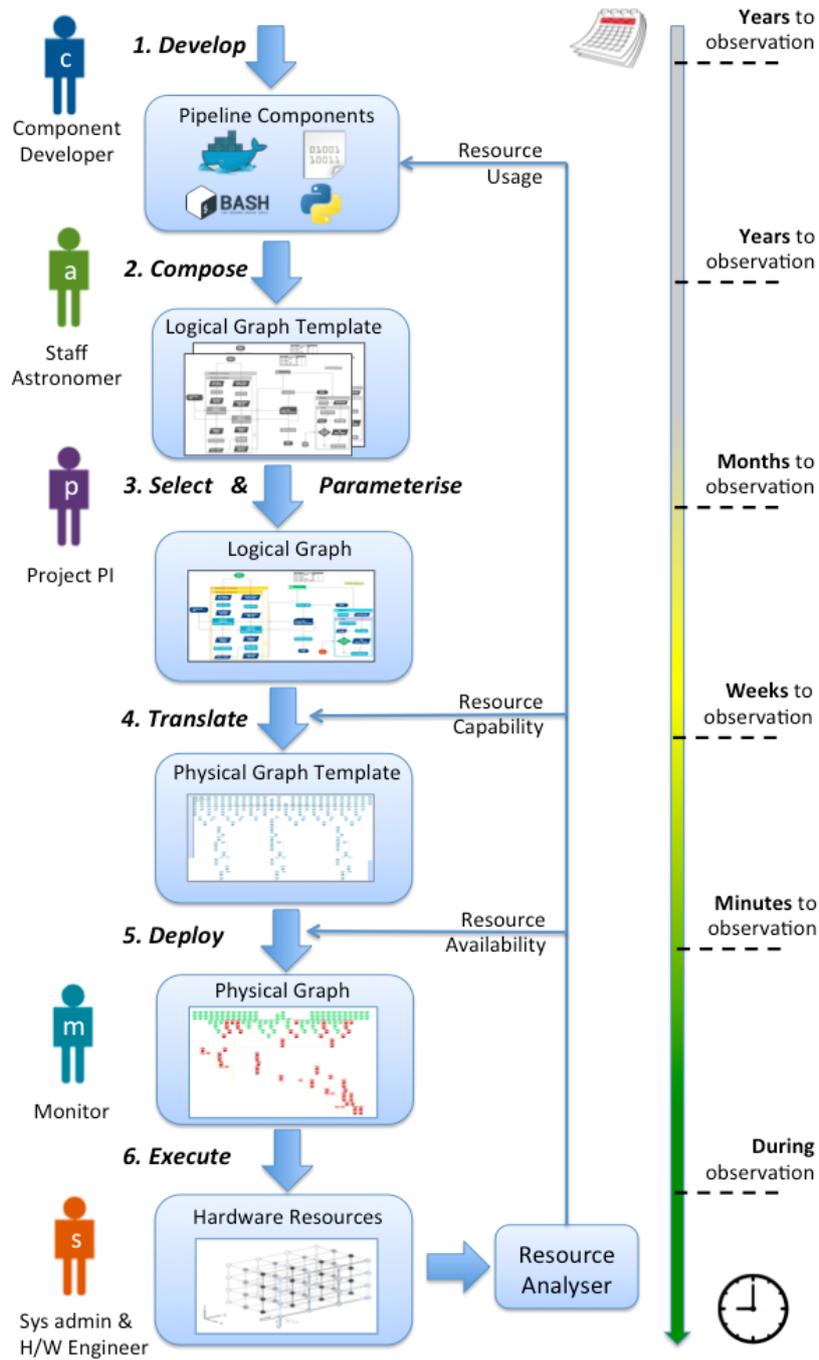

Figure 1: The DALiuGE framework and the different stages of a pipeline's lifecycle in the context of the SKA SDP

**Stage 3.** During the creation of the detailed observation the PI will select one of those released Logical Graph Templates and provide the values for the user specifiable parameters. Together with the observation description this will allow the system to transition the Logical Graph Template to a *Logical Graph*, which is specific for the proposed observation.

**Stage 4.** DALiuGE translates a Logical Graph into a *Physical Graph Template*, an unrolled and logically partitioned version of the Logical Graph not bound to specific physical compute resources. The translation from a Logical Graph to Physical Graph Template is automated and uses *Resource Capability* information obtained from the Resource Analyser in order to achieve certain cost objectives. Note that a single node in the logical graph may correspond to many nodes in a Physical Graph Template.

**Stage 5.** The Deployment stage takes place when an observation or a batch processing job needs to be run. DALiuGE first retrieves the corresponding Physical Graph Template from the graph repository. By correlating the information on *Resource Availability* from the Resource Analyser, which analyses real-time resource usage such as compute node, storage, etc, DALiuGE associates each node in the Physical Graph Template with an available resource unit, effectively generating a *Physical Graph*. Minutes before an observation starts, DALiuGE deploys the Physical Graph onto the *Drop Managers*, a set of daemon processes responsible for the run-time aspect of DALiuGE. This creates *Drops*, the objects responsible of wrapping and managing both the data and the applications contained in the Physical Graph, and of driving the execution of the Physical Graph. A detailed discussion on the concept of Drops will be presented in Section 4.

**Stage 6.** Once the observation finally starts, the Execution stage takes place, and the graph execution cascades down automatically through the graph edges. When all tasks are completed, some data is persistently preserved as science-ready products. DALiuGE Drop managers can be equipped with plugins to any third-party Resource Analysers to continuously profile workload behaviours for each task. Such resource usage information can be fed back to the development of pipeline components.

It is important to note that for a given pipeline the first two stages are executed only once (unless the pipeline logic actually changes, which would produce a new Logical Graph Template) independently of how many observations use the same pipeline definition. The following sub-sections discuss each of these stages in details.

*3.1. Develop*

Figure 2 depicts how pipeline components are developed in a way that separates concerns of four key stakeholders/roles. First, domain specialists focus on designing algorithms to solve a particular domain-specific problem. Algorithm developers implement these algorithms, which are later transformed into ApplicationDrops by the Pipeline component developers. DALiuGE has built-in support for wrapping common task types such as: bash commands, system executables and Docker containers. Pipeline component developers are also responsible for representing and wrapping data sets into Data Drops, either by using an built-in Drop type, extending one, or writing their own.

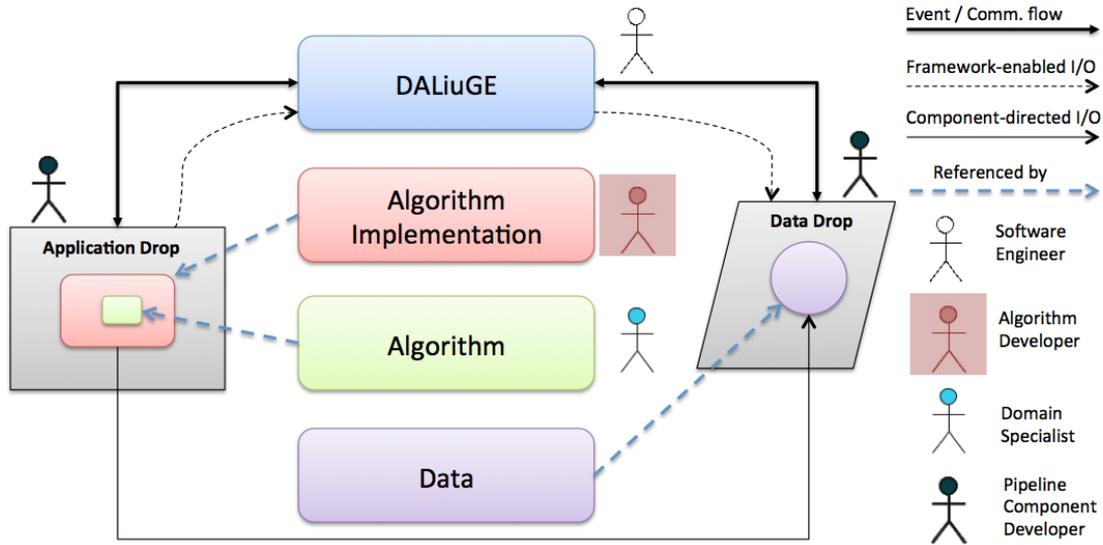

Figure 2: During development time, how Drops are associated with algorithms, pipeline components, data and the DALiuGE graph execution framework which enables separation of concerns.

3.2. Compose

A Logical Graph Template represents logical operations in a processing pipeline without concerning the underlying hardware resources and observation details. The building blocks of a Logical Graph Template are called **Constructs**. Constructs are the elements exposed to the Logical Graph Template developer through the Logical Graph Template editor. The two basic constructs are **Data** and **Component**. They are templates, from which Data Drops and Application Drops are instantiated respectively. Note that one Data or Component construct in a Logical Graph Template could result in the generation of many Data or Application Drop instances in a Physical Graph Template. Each construct has several associated properties, whose values will be populated in the next stage (Section 3.3). In particular, the Component and Data constructs expose the *Execution time* and *Data volume* properties respectively, indicating how long a task should take to run, and how much data is contained in a Data construct. Values of these properties can be directly obtained from parametric models or estimated from the resource usage information given by the Resource Analyser.

Data constructs can be linked to Component constructs as inputs or outputs. This linking rule is vital for designing dataflow-like programs where tasks and data are both nodes of the graph. In addition to the basic constructs, a number of control flow constructs enable the creation of more complex Logical Graph Templates. They form the skeleton of the logical graph, and determine the ultimate structure of the Physical Graph Template to be generated (see Figure 5). DALiuGE currently supports the following flow constructs:

- ***Scatter*** represents data parallelism. Data fed into a *Scatter* construct is broken down into a number of partitions, and the constructs inside a *Scatter* construct consume a single data partition within the enclosing *Scatter*. The `num_of_copies` property of Scatter controls how many partitions should be generated. This is used by DALiuGE when generating parallel branches of execution.

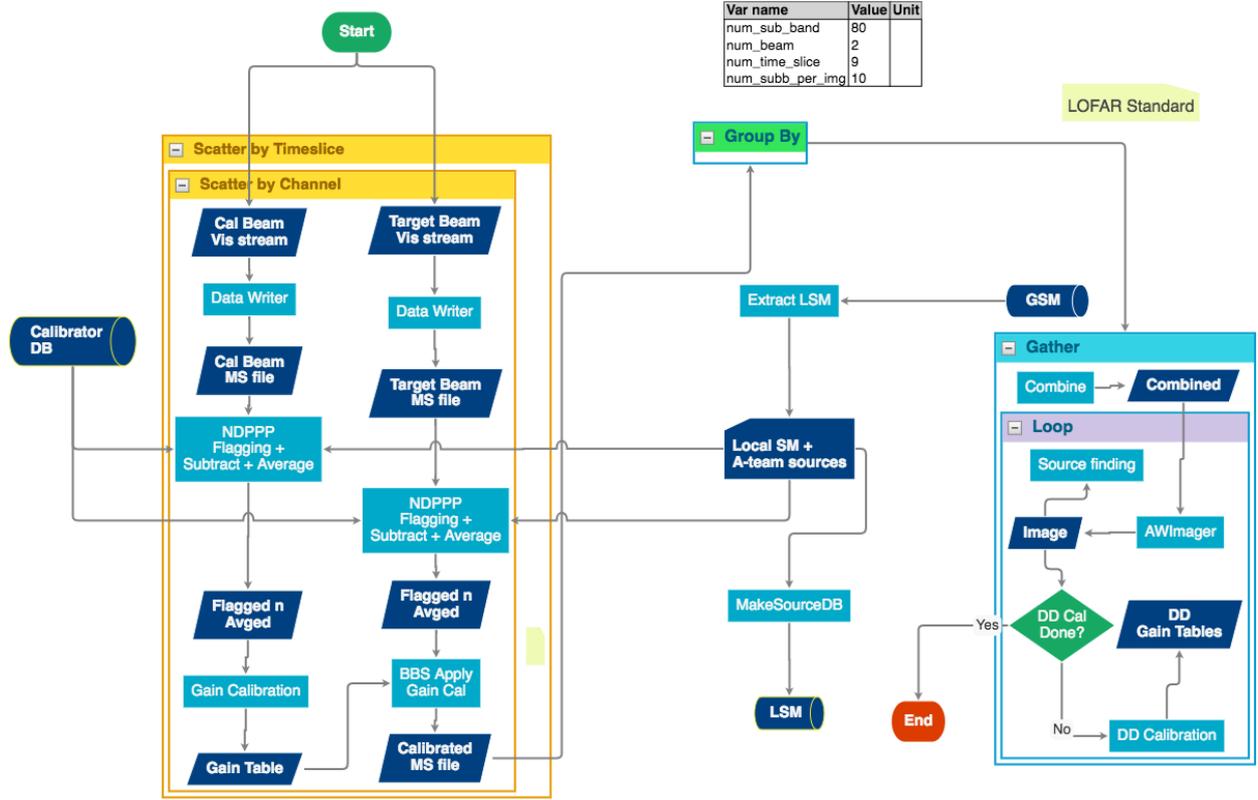

Figure 3: The Logical Graph Template describing the imaging pipeline [47] for the LOFAR telescope [48]. It contains two *Scatter* constructs where the initial data is split by time and frequency domain, a *GroupBy* construct grouping the data by frequency, a *Gather* construct and a *Loop* construct for the imaging process.

- **Gather** represents a data barrier. Constructs inside a *Gather* construct consume a sequence of data partitions as a whole. *Gather* has a `num_of_inputs` property stating how many partitions each *Gather* instance can handle. This in turn is used by DALiuGE to determine the number of *Gather* instances to be generated in the physical graph. *Gather* can be used in conjunction with *GroupBy* (as shown in Figure 3), in which case, data held in a sequence of groups are processed together by components enclosed inside *Gather*.

- **GroupBy** performs data reordering (a.k.a. the corner turning problem [49] in radio astronomy). The semantic is analogous to the `Group By` SQL statement used in relational databases. It is comparable to the "static" MapReduce shuffling, where the keys collected by all `Reducers` are known a priori. Figure 4 shows an example of the *GroupBy* construct. In this example, a list of 2D points were originally sorted based on the axes order of (x, y). After performing *GroupBy*, they are re-sorted based on the axes order of (y, x). DALiuGE requires *GroupBy* be used in conjunction with nested *Scatter* constructs as shown in Figure 3 such that data Drops (e.g. "Calibrated MS file") initially sorted in the order of the outer partition — `Scatter by Timeslice` — are re-sorted in the order of the inner partition — `Scatter by Channel`.

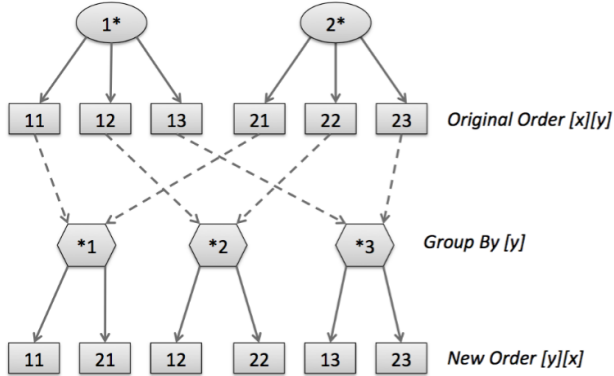

Figure 4: An example of the *GroupBy* construct, which re-sorts a list of 2D points from axes (x, y) to axes (y, x)

- **Loop** indicates iterations. Constructs inside a Loop will be repeatedly produced for a fixed number of times. DALiuGE currently does not support dynamic branch conditions for Loop. Instead, each Loop construct has a property named `num_of_iterations`. An example of the Loop construct is shown in Figure 3.

To let users compose Logical Graph Templates we have developed a Javascript-based Logical Graph Editor, which allows astronomers to use a Web browser to design Logical Graph Templates and manage a Logical Graph Template repository. In an operational observatory environment these astronomers would be trained staff astronomers, familiar with the specifics of the available constructs. The repository is currently implemented as a managed file system directory. Each Logical Graph Template is represented as a JSON-formatted textual file, and can be accessed and modified remotely through the Logical Graph Editor via its RESTful interface.

*3.3. Select & Parametrise*

In order to tighten the text in the following paragraphs and captions we are introducing the following abbreviations:

| | |
|---|---|
| PI | principle investigator |
| OBS | observation |
| LGT | Logical Graph Template |
| LGR | Logical Graph |
| PGT | Physical Graph Template |
| PGR | Physical Graph |
| DM | Drop Manager |

Several months before an observation starts, the science project PI is responsible for generating the Logical Graph for OBS by first selecting an appropriate LGT from the LGT

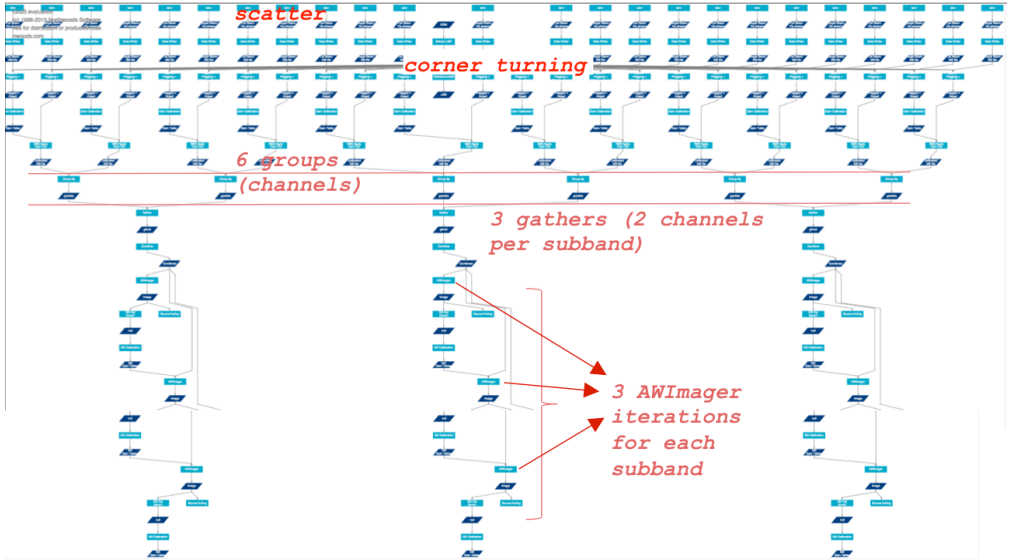

Figure 5: An example of Physical Graph Template unrolled from the Logical Graph Template in Figure 3

Repository and second providing concrete parameter values to all properties in LGT based on the observation schedules. The only difference between LGT and LGR are those parameter values filled in by the project PI.

3.4. Translate

While a Logical Graph provides a compact way to express complex processing logic, it contains high level control flow specifications that are not directly usable by the underlying graph execution engine, and cannot be mapped to resources directly. To achieve that a Logical Graph is translated into Physical Graph Template (PGT). The translation process consists of three steps: validation, construct unrolling and logical partitioning. The first step validates the Logical Graph and its constructs, ensuring that it is in a state in which it can be translated. For example, DALiuGE currently does not allow cycles in the Logical Graph. This procedure is analogous to the syntax checking done by compilers, making sure the structure of the program being processed is correct. The second step unrolls the logical graph by first creating all necessary Drop specifications (which also might include Drops that do not appear in the original logical graph as constructs by themselves), and second establishing directed edges amongst these Drop specifications. This set of drop specifications and the edges linking them together is the PGT. As shown in Figure 5, the PGT example illustrates how logical graph constructs in Figure 3 are unrolled to produce a larger number of Drops based on the constructs semantics.

Finally, the third step divides the PGT into a set of logical partitions and generates an execution sequence of the Drops within each partition such that certain performance requirements (e.g. total completion time, total data movement, etc.) are met under given constraints (e.g. resource footprint, collocation criteria, device locality, etc.). DALiuGE currently assumes hardware resources are homogeneous. Under this assumption, we have imple-mented two sets of graph partitioning algorithms — *min_time* and *min_res*. The *min_time* algorithms produce an optimal number of partitions such that first the total completion time of the pipeline (which depends on both execution time and the cost of data movement on the graph critical path) is minimised, and second at any point in time, the number of drops running in parallel within a single partition is no greater than a Degree of Parallelism (DoP) threshold. The *min_res* algorithms minimise the number of produced partitions subject to satisfying completion deadline and the DoP threshold constraints. Since a partition has a constrained DoP, the number of partitions corresponds to resource footprint. Inspired by the hardware/software co-design method [50] used in embedded systems design, we used look-ahead search strategy as well as several stochastic local search heuristics such as simulated annealing [51] and particle swarm optimisation [52] in order to locate a global optimum during graph partitioning.

*3.5. Deploy*

To execute a PGT DALiuGE first retrieves it from the graph repository, then assigns each Drop specification in the PGT to an available resource unit. In doing so the Physical Graph Template is converted to a **Physical Graph** (PG), which is now bound to actually available hardware at the time of execution. Finally DALiuGE communicates the PG to the **Drop Managers** (DM) for instantiation and execution.

*Resource mapping*

This step maps each logical partition of the PG onto a given set of currently available resources in certain optimal ways (load balancing, minimum cost of data movement, etc.). Here we have adopted a two-phase scheduling approach [53] to separate the previous graph partitioning step, which is somewhat oblivious of underlying resources, from the resource mapping step, which assigns each Drop onto a physical compute node in the cluster. Such placement requires real-time information on resource availability from the **Resource Analyser** as shown in Figure 1. Similar to the last step, we currently assume resource pools consisting of nodes with identical capabilities of computing, storage, and inter-connect. We use the METIS software library [54], which internally uses a multilevel k-way partitioning algorithm [55], to merge the $p$ PGT partitions into $m$ virtual clusters if $p > m$, where $m$ is the number of currently available machines with the goal of balancing the overall workload (both compute time and memory usage) evenly. The physical mapping from the $m$ merged clusters to $m$ compute nodes becomes a straightforward round-robin assignment if the resources are all homogeneous. Once a mapping is established from a cluster $c$ to a compute node $n$, all Drops in the cluster $c$ will be assigned a network address of compute node $n$.

*Drop Managers*

**Drop Managers** are the set of daemon process that manage and execute PGs. DMs offer a unified interface to external users to interact with the run-time system, allowing users to submit and deploy physical graphs and to query and monitor graph execution status. Drop Managers are organised hierarchically for scalability considerations. The hierarchy levels currently implemented include three levels, but the current design is flexible and allows adding more intermediate levels if necessary in the future. A **Node Drop Manager** exists for each compute node in the system and sit at the bottom of the DM hierarchy. They are ultimately responsible for creating and deleting Drops. Because compute nodes are grouped into Data Islands, a **Data Island Drop Manager** exists at the Data Island level, managing a list of Node Drop Managers. Finally, in order to expose a single point of contact a **Master Drop Manager** manages all Data Island Managers.

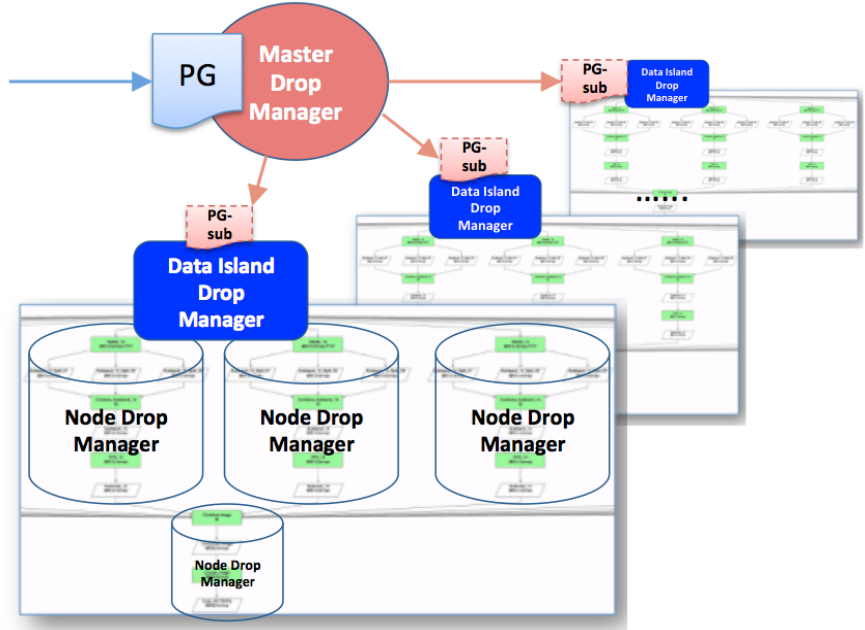

Figure 6: Physical graph deployment across the Drop Manager hierarchy

Drop Managers introduce the concept of a **Session** to represent a given PG execution. Sessions are completely isolated from one another. This enables multiple PGs to be deployed and to executed in parallel within a given Drop Manager as long as the resource availability suggested by the Resource Analyser is sufficient enough to produce valid resource mappings during the translation step. Sessions have a simple lifecycle: they are first created, then a complete or a partial PG is attached to them, after which the graph can be deployed. This leaves the session in a running state until the graph has finished its execution, at which point the session is finished and can be deleted.

*Physical Graph Deployment*

Once a Drop Manager receives a PG it prepares the graph deployment on their managed resources as prescribed in the PG. The deployment process recursively traverses the DM hierarchy as shown in Figure 6. For Node Drop Managers (the lowest level in the DM hierarchy) this involves checking the validity of the PG and the creation of the Session and all its Drops. For a Data Island Drop Manager $d$ this process involves separating the PG based on the node placement information. This way Drop specifications belonging to the same compute node $n$ form a PG sub-graph with no edges pointing to Drop specifications

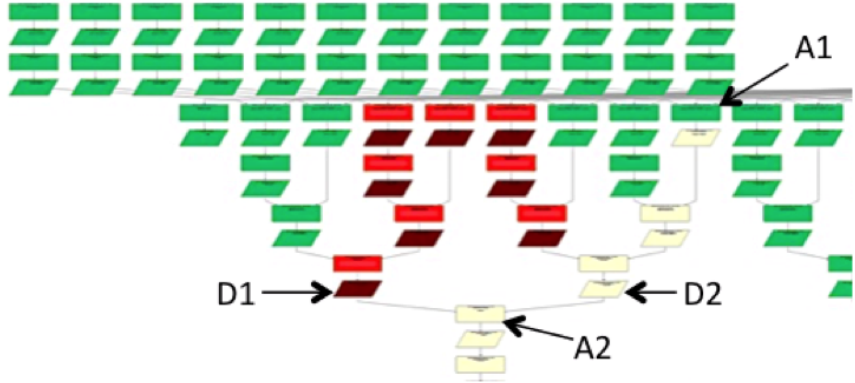

Figure 7: A Physical Graph execution in case of failure. The Application Drop *A2* is configured with $t = 50\%$ and therefore can still continue its execution if the Data Drop *D2* is successfully completed

in other nodes. $d$ then submits the sub-graph to the Drop Manager running on $n$, which is a member of $d$. The edges from the original PG crossing node boundaries are recorded by the Data Island Drop Manager and communicated later to the relevant Node Managers for creating the necessary connections between them. The Master Drop Manager splits the PG based on island placement information following the same recipe.

*3.6. Execute*

During the execution phase the root Drops of the Physical Graph (i.e., those without inputs or producers) are triggered to start their execution. In the case of root Data Drops, their data is considered to be present and therefore they are marked as completed. Once a data Drop is marked as complete it will fire an event to all its consumers. Consumers will then decide if they can start their execution depending on their nature and configuration. Batch-style Application Drops in particular wait until all their inputs are completed to start their execution. On the other hand, data Drops receive an event every time their producers finish their execution. Once all the producers of a Drop have finished, the Drop marks itself as complete, notifying its consumers, and so on. This way drops have the ability to drive their own execution instead of relying on a central orchestrating entity.

Failures on Drops are propagated in a similar fashion automatically via events. Data Drops move to the ERROR state if any of its producers move to ERROR. An Application Drop moves to ERROR if an error-tolerant threshold, $t$, inputs have moved to ERROR. Setting $t$ to a value greater than zero allows certain branches of execution to fail without preventing the main execution branch from making progress if enough inputs are present after reaching an error-tolerant gathering point.

Figure 7 shows a physical graph execution with simulated failures. To produce this execution we have intentionally let some Application Drops fail (either raising random exceptions or blocking the event flow) at the upstream of the graph execution. Consequently many downstream Drops have been marked in Red as failure through event propagation. Since the error-tolerant threshold $t$ was set to 50% in this example, the light-yellow Application

Drop *A2* is still waiting for one of its inputs *D2* to get ready although the other input Data Drop *D1* has already entered the ERROR state. The root cause of such waiting is the artificially blocked event flow from the Application Drop *A1*. As a result many downstream Drops are also in the WAITING state. Eventually *A2* will also enter ERROR due to timeout, rendering the entire rest of the main branch in the failure mode.

*3.7. Implementation details*

DALiuGE is currently implemented in Python and publicly available on GitHub[3]. Supported Python versions are 2.7, 3.3, 3.4 and 3.5. Continuous integration (i.e., building and testing) is available via Travis-CI[4]. Installation of the framework is based on `setuptools` and therefore follows the standard installation procedure for python packages (via `pip` or similar tools), automatically retrieving and installing all requirements. DALiuGE has built-in support for common storage platforms (filesystem, in-memory, S3) and applications (scp, `bash` commands, Docker, TCP socket listening, etc.).

Drop events flowing through different nodes travel via ZeroMQ PUB/SUB sockets using pyzmq [56]. These sockets are set up by the Node Managers using the edges information of the Physical Graph communicated by the higher levels of the Drop Management hierarchy. Other forms of Remote Procedure Calls used by Drops to interact with each other are implemented via ZeroRPC [57], which is built on top of pyzmq, but other libraries like Pyro4 and RPyC are also supported and available via environment variable settings. Drop Managers expose a simple HTTP REST interface for monitoring and control.

We currently use JSON as the serialization format for the different graphs. JSON-encoded graphs are compressed and uncompressed on-the-fly when transmitted. We parse the JSON content iteratively to keep memory low for big graphs using a modified version of the ijson library. The modifications achieved a $\sim$10x performance increase, making it comparable to the built-in `json` Python module.

*3.8. Framework overhead*

This section briefly discusses the framework overhead introduced by DALiuGE. We present the detailed analysis on DALiuGE scalability and framework overhead in our companion paper [13]. .

Figure 8 shows that the framework execution overhead when increasing the number of Drops up to 12.6 million while keeping the number of nodes to 400. We make two observations from Figure 8. First, although the execution overhead per Drop is relatively small (below 10 microseconds) for both cases, in all four graphs, the overhead for 5 islands is less than 5 microseconds, almost half of that in the single island case. This is expected since the run-time management overhead is now distributed to multiple island managers. Second, initially each one of the 400 nodes still has idling resources to accommodate more Drops for graph *A* with 2.1 million Drops. Graph *B* with over 4.2 million Drops has a slightly smaller overhead than Graph *A*. But as we double the number of Drops to over 8 million, the overhead stays flat (for 5 islands) or goes up (for single island) with less idling resources available per node. In addition to the execution overhead per Drop there are also setup and deployment overheads, both of which are essentially constant for a given environment and graph and they are also not affecting the overall execution time directly, since they occur on different time scales. In particular the setup overhead will occur just once during the initialisation of the environment, when the various DALiuGE daemons are started on the compute nodes. After that DALiuGE can execute many graphs. In an operational environment the initialisation is expected to happen only quite rarely and the overhead is thus negligible for a single graph.

---

[3]https://github.com/SKA-ScienceDataProcessor/dfms
[4]https://travis-ci.org/SKA-ScienceDataProcessor/dfms

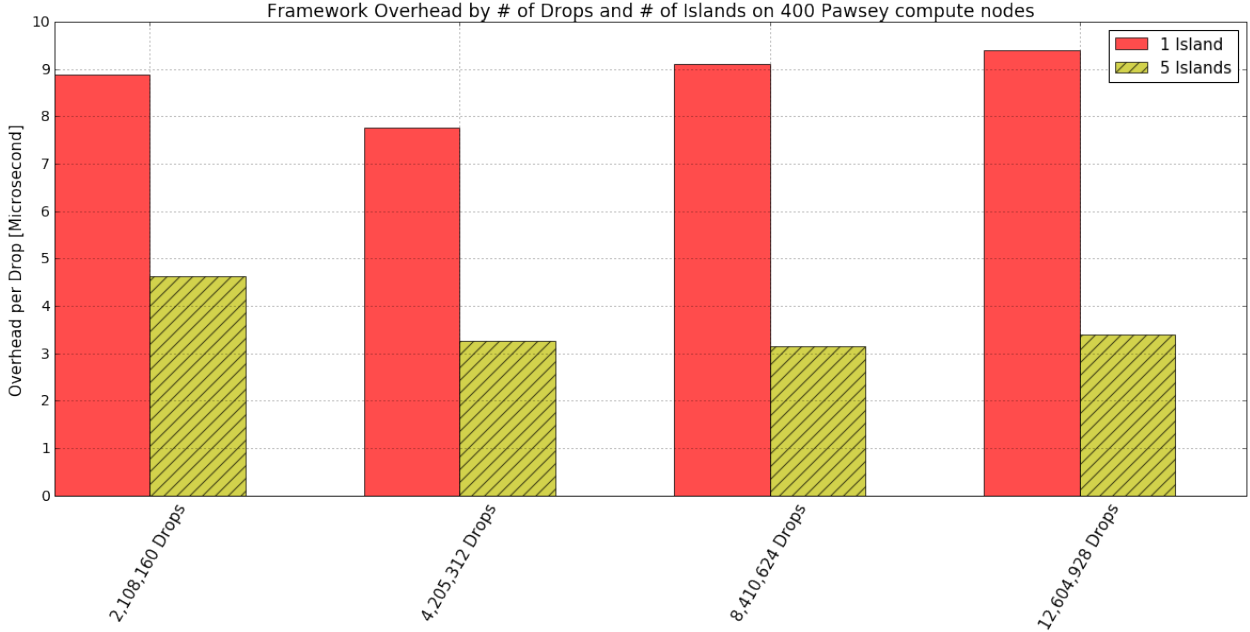

Figure 8: framework overhead on 1 and 5 islands on 400 compute nodes. Y-axis is the overhead measured in microsecond per Drop

## 4. Drop

In a nutshell, Drops [58] are software objects wrapping a generic payload. Figure 9 shows the simplified UML diagram of the Drop system in DALiuGE. All the various Drop types, described in the next subsection are derived from the abstract `Drop` class. The abstract `ContainerDrop` class implements a hierarchical grouping of Drops into a single entity, and currently has two concrete sub-classes — `ApplicationDrop` and `DataProductDrop`. The `ApplicationDrop` container allows to group together everything required to make an actual application stand-alone into a single Drop. This is equivalent to the way applications are bundled in Mac OSX [59] or inside a Docker container [46]. The `DataProductDrop` container allows to bundle Drops together, which belong to a specific SKA data product. In general such a `DataProductDrop` will consist of thousands of individual Drops, including

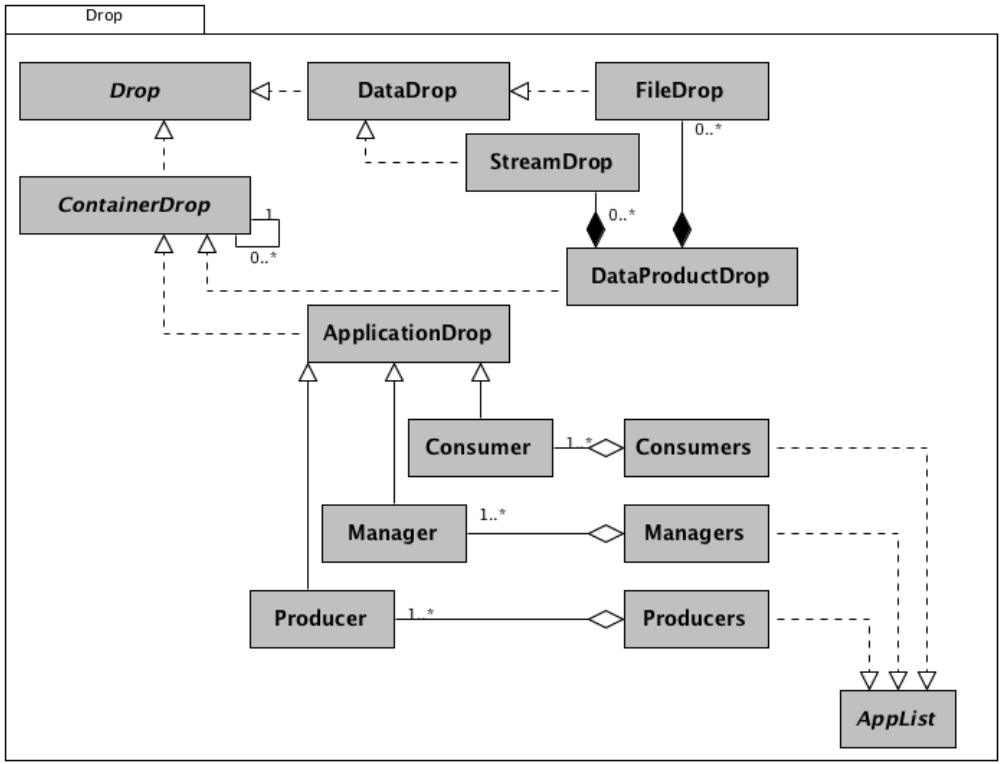

Figure 9: The Drop object diagram in UML notation. This diagram shows the abstract Drop object as well as some of the existing concrete objects.

Data Drops and potentially ApplicationDrops, depending on the definition of what the actual data product entails. For the system this design allows to manage and handle data products as single entities.

The Drop object wrapper associates methods and provenance, as well as life-cycle and access control data with the payload, thus making data virtually active. In particular it means that a Data Drop can react and raise events, which then drive the actual execution of the Physical Graph in DALiuGE. While a Drop's payload is stateless (write once and read multiple times), Drops themselves are stateful, which not only allows us to manage Drops through persistent check-pointing, versioning and recovery after restart, but also enables data sharing amongst multiple processing pipelines in situations like re-processing or commensal observations. Each Drop exposes its payload location via a URL, has an unique identifier, and is instantiated, monitored and destroyed by a Drop Manager, which subscribes to the events fired by the Drops as they transition through their different states.

Figure 10 shows how Application Drops and Data Drops are connected to each other with the former being either *consumers* or *producers* of the latter (or *inputs* and *outputs* from the opposite point of view, respectively). DALiuGE distinguishes between normal inputs/consumers and their streaming counterparts. A normal consumer application works in a batch-like fashion, running only after all its inputs are marked as completed, which in turn occurs only when their producer applications have finished. Streaming consumers work instead by continuously consuming their input data, which is continuously produced in turn by a producer application.

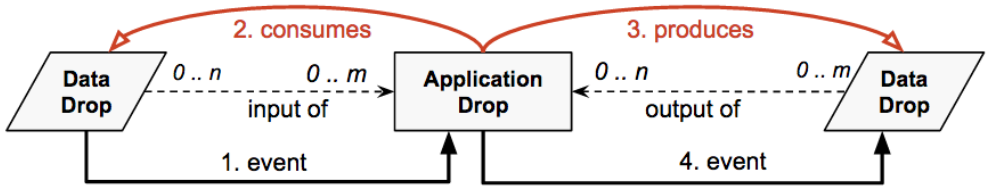

Figure 10: Data drops and application drops are conceptually related and connected to form an event flow/graph. Note that there is a n..m relationship between Data Drops and Application Drops, allowing for scattering, gathering and re-ordering applications.

### 4.1. Drop Channels

During their execution, Drops need two different channels of communication. The first is to fire events to interested parties, which is a crucial mechanism for the graph execution as explained in §3.6. The second is to be able to interact with other Drops and query information about their properties, such as their `dataURL`, `uid` or other properties. In both cases, Drops residing in the same node (and therefore in the same process as the Node Manager) use direct object invocation to achieve communication. If two Drops reside in two different nodes we use a publish/subscribe mechanism to fire events between nodes, and a Remote Procedure Call interface to allow Drops to communicate with each other. Note that these channels are merely communication channels, they do not carry the data payload. The DALiuGE design very cleanly separates communication from bulk data operations, which allows us to change the underlying mechanisms independently, but it also allows to bind those two very different traffic types to dedicated networks, if the underlying platform provides that distinction like for instance the former Fornax supercomputer at the University of Western Australia [60].

### 4.2. Drop I/O

Currently two options are available for Application Drops to perform I/O operations on Data Drop payloads:

1. DALiuGE Data Drops expose their data payload via a simple I/O abstraction layer. This is designed based on the basic `open`, `read`, `write` and `close` POSIX calls, which follow a byte stream data model. A Pipeline component developer can use these built-in POSIX calls ("Framework-enabled I/O" dotted arrows in Figure 2). Moreover, the abstraction can be extended by refining a set of reusable classes, which allow I/O framework developers to realise more sophisticated I/O methods or optimisations (e.g. RDMA-based data transfer). In either case, the data is passing through the Drop instance, and the I/O framework will take care of remote I/O access across node boundaries. Neither the pipeline component developers and even less the logical graph developers will be exposed to those details. In particular this enables a fully transparent replacement of both storage methods and as well as storage formats in order to achieve a better throughput.

2. The pipeline component developer can instruct the algorithm implementation to directly perform I/O without going through DALiuGE ("Component-directed" solid arrows in Figure 2). In this case, the Application Drop obtains the `dataURL` (e.g. POSIX file paths, RDMA buffer pointers, etc.) of the Data Drop, and passes it onto the algorithm implementation as a parameter. It is the responsibility of the pipeline component developer to ensure that I/O is occurring in the correct location.

*4.3. Drop Lifecycle*

Drops follow a basic lifecycle consisting of several states as shown in Figure 11. A Drop (Data or Application) starts in the INITIALIZED state, indicating its payload is not fully present. For Data Drop, this means its enclosing data content is not available. For example, the file owned by a `FileDrop` has not yet been created on the designated file system. For Application Drop, this simply means it is not executable yet. For example, it takes time to serialise and deserialise a "remote" function[5] across compute nodes within a network. From INITIALIZED a Drop becomes COMPLETED once its payload is fully written, optionally passing through WRITING if the payload is being written through the DALiuGE I/O framework. Once in the COMPLETED state the payload can be read / executed as many times as needed. Eventually Drops transition to an EXPIRED state based on a lifetime defined at creation time, after which they deny any further reads. After expiration the data is deleted and the Drop moves to the final DELETED state. If any I/O error occurs the Drop moves to the ERROR state.

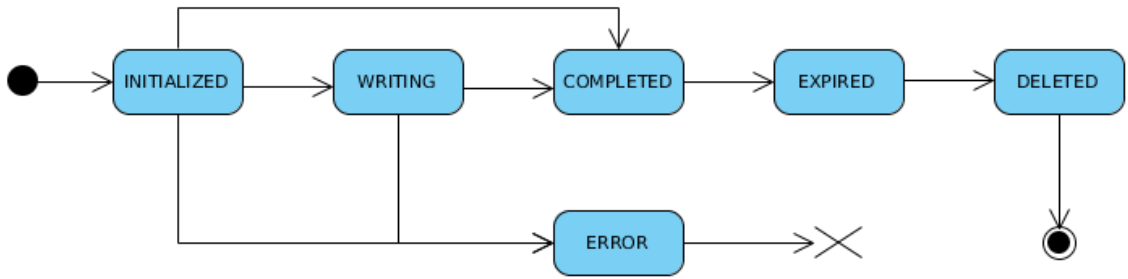

Figure 11: A Drop's lifecycle consists of several states applicable to both Data Drops and Application Drops.

## 5. Case Study I — CHILES on AWS

The Cosmos HI Large Extragalactic Survey (CHILES) [61] is running at the Karl G. Jansky Very Large Array (VLA), a 27 antenna array. The new, upgraded front-end (wideband L-band receivers) and back-end (the WIDAR correlator) [62] provide instantaneous

---

[5]Equivalent to copying a JAR file wrapping Hadoop/Spark tasks to remote JVMs/Executors

coverage for spectral line observing between ∼940 and ∼1430-MHz on the sky (15 spectral windows of 32MHz, each giving a total of 480MHz in each session). The array configuration for CHILES is VLA-B, which has 11 km baselines and a typical beam size of ∼5x7 arcsec at 1.4 GHz, assuming natural weighting. The observations are dithered in frequency to smooth out the edges of the spectral windows. The antennas (being 25m in diameter) see about 0.5 degrees (∼2000 arcsec) across the sky at the pointing centre. The project forms images with 2048 pixels of 1 arcsecond in size during this development phase. This data is in 15.625kHz channels, contains 351 baselines, a little less than 31,000 channels per polarisation product to be processed, and a full field of view of 2048x2048 pixels in the image plane.

In our previous work on CHILES data reduction [20], we had to hard code scripts for three different computing environments. Therefore, we decided to use DALiuGE to drive the execution of the "new" pipeline, which included 5 pipeline components:

1. Splitting the measurement set for an individual day into measurement sets containing 4Mhz chunks. The current implementation has 42 days worth of data and this will grow with the release of the next semester's data.
2. Subtracting the local sky model and moving the corrected fluxes into the "Corrected" column of the measurement set on each of the 4 MHz chunks across all days.
3. Cleaning by taking all the days measurement sets for a 4MHz band and apply clean to them all to produce a single cleaned file for that 4MHz band.
4. Converting each 4MHz measurement set to JPEG2000.
5. Concatenating the final cleaned 4Mhz files into a single measurement set.

5.1. Setup

Setting up AWS for this exercise consisted of three simple steps. Step 1 chooses the correct AWS EC2 instance to match the CPU, memory, disk and I/O bandwidth requirements of the task. Step 2 runs a Python script to start the required number of DALiuGE Drop Manager servers, ensures that disks are initialised, and the correct software is installed; and starts a data island manager to orchestrate the deployment of the physical graph to the various Drop Managers. Step 3 runs a Python script that calculates what is required to be done and submits the physical graph to the data island manager.

As mentioned in §3.7 the installation of DALiuGE on AWS was very straightforward. To speed up the start up times of the EC2 instances we created an Amazon Machine Image (AMI) with a recent version of DALiuGE pre-installed and a python virtual environment pre-created ready to be used. At instance-creation time we pulled the latest version of DALiuGE from GitHub and installed it into the pre-created virtual environment to make sure we were working with the latest version.

The installation of the processing components was done primarily using Docker. The CHILES data is processed via a number of CasaPy [63] tasks and SkuareView [64]. We wrapped the CasaPy and SkuareView code inside Docker images and uploaded them to DockerHub. By doing this we did not need to compile or install any additional software other than the Docker daemon within our EC2 instances. DALiuGE has built-in Docker support, and therefore very little effort was needed to integrate the processing components into the pipeline graph.

When an Application Drop is triggered to "run" it first executes a Python script using boto3 [65] to set up a cluster of EC2 instances. As the EC2 instances start they launch the corresponding Drop Managers. In the mean time the corresponding Physical Graph is calculated, and once the Drop Managers are up and ready the Physical Graph is deployed onto DALiuGE and executed. The Physical Graph includes a special Application Drop that automatically shuts down each of the EC2 instances, making sure we do not get charged for servers that are effectively doing nothing. For cost reduction, we request Amazon EC2 Spot Instances (a bid-driven process that makes instances available at discount prices of 40-80% when compared to the on-demand price) rather than directly launch on-demand ones. The Python script always looks for the cheapest spot price in the regions specified for the instance type required depending on the load characteristics of the pipeline.

The configuration of each EC2 instance is controlled at creation time by submitting a YAML file to configure the packages to be loaded, and the AWS and Docker files to be created; and a Bash file to configure the disks, pull the latest version of docker containers from DockerHub, pull the latest code for DALiuGE and the CHILES pipeline from Github, and start the DALiuGE Drop Manager. We used three AWS storage options during different stages of the pipeline — Elastic Block Store (EBS), Ephemeral Storage, and Simple Storage Service (S3).

## 5.2. Results and Costs

The initial measurement sets for semester 1 of the CHILES project amounts to approximately 15TB of data[6]. Splitting the original measurement sets into 4MHz chunks creates 120 sub directories each contain 42 measurement sets which requires 9.1TB. The model subtraction requires 9.1TB. The clean step creates 120 4GB files for a total of 480GB of data. The final concatenation produces a single file of 480GB.

The `splitting` process requires either an i2.2xlarge or an i2.4xlarge EC2 instance depending on the size of the measurement set on a particular day. If the measurement set is greater than 500GB we required the greater power (and more expensive) i2.4xlarge instance. The `model subtraction` requires an i2.2xlarge EC2 instance. It takes approximately 18 hours to perform the 5,040 model subtractions if given 20 nodes. The spot price for an i2.2xlarge is usually of the order of US$0.20 per hour which gives an indicative cost of US$72. The `clean` process requires an i2.4xlarge EC2 instance because of the increased load on the CPU and memory. it takes approximately 20 hours to perform the 120 cleans if given 20 nodes. The spot price for an i2.4xlarge is usually of the order of US$0.35 per hour which gives an indicative cost of US$140. The `conversion` to the JPEG2000 images is performed using a single i2.2xlarge instance and only takes a few hours. The final `concatenation` has only been performed once as the resultant image is too large to be downloaded quickly and for systems to process. It took over 48 hours using a i2.4xlarge instance.

---

[6]Data transfer from non-AWS sources into AWS EC2 instances is free of charge

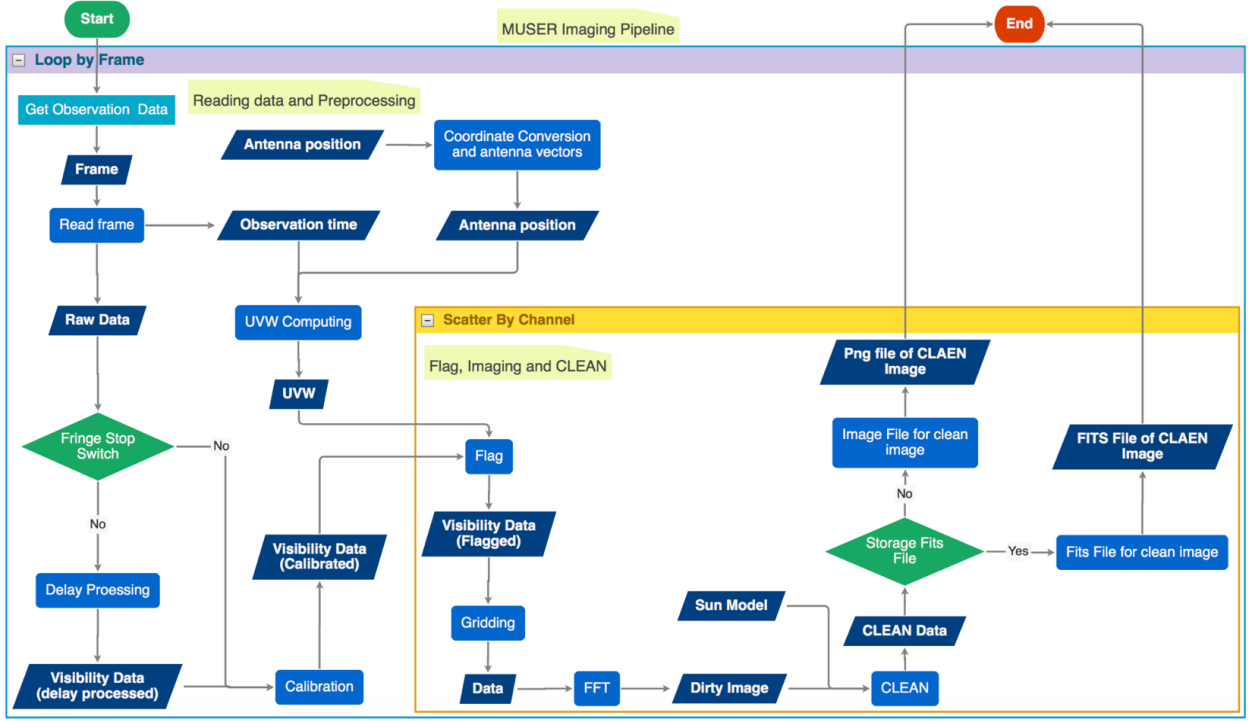

Figure 12: The Logical Graph of the MUSER pipeline

## 6. Case Study II — MUSER

The Mingantu Ultrawide Spectral Radioheliograph (MUSER) [66] is a solar dedicated synthetic-aperture radio interferometer that is capable of observing radio bursts and producing high-quality radio images at frequencies from 400MHz to 15 GHz with high temporal, spatial, and spectral resolution. It consists of 100 radio antennas spirally distributed in Ming'antu, Inner Mongolia, China. The RF signal of MUSER in 0.4–15 GHz is divided into 40 antennas of 4.5 m at 0.4–2 GHz (subarray MUSER-I), and 60 antennas of 2 m at 2–15 GHz (subarray MUSER-II) bands.

To design and develop the data processing pipeline of the MUSER, a novel distributed-computing infrastructure, OpenCluster [67], was previously implemented in Python. To evaluate the usability and performance of DALiuGE, we have "migrated" all pipeline components under the OpenCluster to Drops. We created 12 pipeline components such as raw data acquisition, frame data distribution, dirty image processing, CLEAN, and so on as shown in Figure 12.

Based on the original design of the digital correlator, the output data are encapsulated into a data frame that has 16 frequency channels. We effectively distributed the frame data into different processes by using the DALiuGE *Scatter* construct. Eight servers with a total of 64 CPU cores were deployed to run the pipeline using the DALiuGE framework. Although all communications are based on the 10G Ethernet network, we carefully considered the design of Data Drops to ensure the pipeline will not be bounded by the I/O sub-system, which is largely determined by Data Drops. In particular, we used DALiuGE built-in `InMemoryDataDROP`s to store visibility data and ephemeris data because data of these types needs high I/O bandwidth to support continuous data processing. The `FileDROP` was used in less I/O intensive situations such as the data archive.

In our preliminary experiment and test, we obtained satisfactory science results and the data processing performance is comparable to the OpenCluster-based pipeline. More importantly, we found the new pipeline based on DALiuGE can be easily developed by re-using existing, mature pipeline software modules. This capability of "enabling composition of existing processing units" is one of the main factors to drive the SDP architecture [10].

## 7. Conclusion and Future Work

In this paper, we provided the technical overview of DALiuGE, an execution framework developed to meet the challenges imposed by the astronomical data deluge in general [68, 69, 70], and by the SDP element of the SKA in particular. At its core, DALiuGE is a flexible yet scalable graph execution engine that can be used in a wide variety of platforms including both supercomputers and cloud environments.

DALiuGE represents large-scale, distributed processing pipelines using graphical models. Such a graph representation allows us to deterministically exploit as many parallelism opportunities as possible, which leads to a completely decentralised, scalable execution engine. Moreover, graph modelling enables us to tap into techniques and algorithms well developed in Graph theory and Combinatorics. This helped us to solve constrained optimisation problems such as calculating optimal execution plans from logical pipeline definitions given resource capabilities and availability. The benefits of *Separation of Concerns* is harvested through the use of graph models in DALiuGE.

Our performance evaluation showed that DALiuGE introduces very little overhead to the pipeline execution latency, scaling up graph deployment smoothly from a single node to thousands[7] of compute nodes without encountering any major bottlenecks. Its implementation maturity was further demonstrated by two real-world case studies, in which DALiuGE has been used for driving the data reduction pipeline in production. DALiuGE is still a work in progress. For future work, we will focus mainly on four areas.

*Logical Graph Template development*

The current prototype of the Logical Graph Template editor is missing a number of crucial features and shows general usability issues. Since this editor is the main visible part of the DALiuGE system it needs quite a bit more attention in the future. There are several advanced open and closed source options, which we need to evaluate in much more detail. One is the currently used JavaScript library GoJS. This library has quite a number of additional features, which we have not fully exploited, yet. Other options include the Google Blockly library [71], the excellent stand-alone application Quartz Composer from Apple [72] and the IoT open source Javascript editor Node-RED [73].

---

[7]The largest DALiuGE deployment thus far consisted of 1300 Magnus compute nodes

*Graph handling*

Graph partitioning is an on-going research area for DALiuGE, and we plan to remove the uniform resource assumption and integrate probabilistic mapping algorithms [74] to deal with heterogeneous resources, particularly the difference in network communication cost between pairs of compute nodes. Currently the initial physical graph submitted to the Drop manager is fully loaded into memory before any further processing such as splitting, redistributing, etc. To completely remove the potential bottleneck imposed by the memory capacity for deployment of large graphs with hundreds of millions of Drops, we are investigating incremental graph unrolling, loading, splitting, redistribution and the use of a declarative representation to express Physical Graphs for such incremental graph processing.

*Failure handling*

We envisage that an on-going work area for large-scale, distributed systems like DALiuGE, is failure handling. In particular we are interested in handling node-level failures by dynamically migrating Drops from failed nodes to healthy ones (if deemed necessary and feasible by the LMC and the Resource Analyser) in order to resume their execution there. To achieve this, and to retain their state information, we are experimenting with a master-to-master replication library called CEDA [75] developed by *ThinkBottomUp*. It uses distributed Operational Transform algorithms [76] to achieve eventual state consistency between replicated Drops, which allows DALiuGE to re-construct them elsewhere in response to failures.

*Advanced I/O handling*

Currently DALiuGE provides a simple POSIX-compliant I/O abstraction layer with built-in support for a few storage media. We are investigating the integration with alternative I/O frameworks to handle intensive I/O activities and large volumes of data communication between Drops in an intelligent and locality-aware fashion. SHORE [77] is such a framework that allows user applications to access various storage hierarchies and backends through a unified application interface. In particular, DALiuGE can utilise SHORE's database to physically locate the data payload associated with a given data Drop, facilitating inter-node data streaming and transport in cases where in-situ processing is not applicable.

# 8. Acknowledgement

The authors thank two anonymous reviewers for their valuable comment in improving this paper. The authors thank Ger van Diepen from ASTRON for suggestions on developing the logical graph template for the LOFAR imaging pipeline.

The International Centre for Radio Astronomy Research (ICRAR) is a joint venture between Curtin University and The University of Western Australia with support and funding from the State Government of Western Australia. This work was supported by resources provided by the Pawsey Supercomputing Centre with funding from the Australian Government and the Government of Western Australia. The CHILES data reduction work was supported by grants from Amazon Web Services and the AstroCompute project. The authors acknowledge the help from the National Supercomputer Center in Guangzhou. This work was supported by the National Natural Science Foundation of China (NSFC) No. 61221491,U1231205,U1631129 and the Open Fund from HPCL No.201401-01.